\documentclass[%
 reprint, 
 amsmath,amssymb,
 aps,
pra,
floatfix,
]{revtex4-1}
\usepackage[normalem]{ulem}
\usepackage{amssymb}
\usepackage{graphicx}
\usepackage{epsfig} 
\usepackage{dcolumn}
\usepackage{bm}
\usepackage{epstopdf}
\usepackage{multirow}
\usepackage{amsmath}
\usepackage{booktabs}
\usepackage{color}
\usepackage{gensymb}
\usepackage{dcolumn}
\usepackage{kantlipsum}  
\usepackage{hyperref}
\usepackage{ulem}
\usepackage{braket}
\usepackage{physics}
\usepackage{comment}
\usepackage[utf8]{inputenc}
\usepackage[T1]{fontenc}
\usepackage{mathptmx}
\usepackage{natbib}

\newcommand*{\mr}{\mathrm} 
\newcolumntype{.}{D{.}{.}{-1}}
\begin{document}

\title{Overlapping Bose-Einstein Condensates of $^{23}$Na and $^{133}$Cs}

\preprint{APS/123-QED}

\author{Claire~Warner}
\author{Aden Z.~Lam}
\author{Niccol{\`o} Bigagli}
\author{Henry~C.~Liu}
\author{Ian~Stevenson}
\author{Sebastian~Will}\email{Corresponding author: sebastian.will@columbia.edu}
\affiliation{%
  $^1$Department of Physics, Columbia University, New York, New York 10027, USA 
}

\date{\today}


\begin{abstract}
We report on the creation of dual-species Bose-Einstein condensates (BECs) of $^{23}$Na atoms and $^{133}$Cs atoms. We demonstrate sympathetic cooling of Cs with Na in a magnetic quadrupole trap and a crossed optical dipole trap, leading to Na BECs with $8(1) \times 10^5$ atoms and Cs BECs with $3.5(1) \times 10^4$ atoms. Investigating cross-thermalization and lifetimes of the mixture, we find that the Na and Cs BECs are miscible and overlapping, interacting with a moderate interspecies scattering length of $18(4)\,a_0$ at $23\,$G and $29(4)\,a_0$ at $894\,$G and coexisting for tens of seconds. Overlapping condensates of Na and Cs offer new possibilities for many-body physics with ultracold bosonic mixtures and constitute an ideal starting point for the creation of ultracold ensembles of NaCs ground state molecules.
\end{abstract}

\maketitle

\section{Introduction}
Mixtures of ultracold atoms play a central role in the creation of complex many-body systems with ultracold quantum gases. They have opened the door to studies of coupled  superfluids~\cite{Barbut2014}, polarons~\cite{Schirotzek2009,spethmann2012dynamics, Hu2016, Jorgensen2016}, Efimov physics~\cite{bloom2013tests, pires2014observation, tung2014geometric}, and impurity physics in optical lattices~\cite{SOspelkaus2006,Will2011}. In addition, ultracold atomic mixtures have enabled the efficient preparation of quantum gases via sympathetic cooling, which has been demonstrated for bosons~\cite{myatt1997production,modugno2001bose} and fermions~\cite{Truscott2001,hadzibabic2002two} and has become the stepping-stone for foundational studies of ultracold Fermi gases~\cite{ketterle2007}. More recently, the assembly of ultracold gases of dipolar molecules~\cite{ni2008high, molony2014creation, park2015ultracold, guo2016creation, takekoshi2014ultracold} has driven the exploration of novel atomic mixtures, including mixtures of alkalis with alkaline-earth~\cite{pasquiou2013quantum, wilson2021quantum} and lanthanide atoms~\cite{hara2011quantum, hansen2011quantum}.

Among the bialkali mixtures, Na and Cs is the only combination for which simultaneous quantum degeneracy has not been reported yet. This is somewhat surprising as individual BECs of $^{23}$Na~\cite{davis1995bose} and $^{133}$Cs~\cite{weber2003bose} have long been investigated, as well as mixtures of each with other  species~\cite{hadzibabic2002two, park2012quantum,wang2015double, mudrich2002sympathetic, grobner2016new, lercher2011production, mccarron2011dual}.  
Studies of Na-Cs mixtures have so far focused on laser-cooled ensembles \cite{shaffer1999trap}, which enabled  photo-association of molecules in a magneto-optical trap~\cite{shaffer1999photoassociative, Kleinert2007, Haimberger2009}, and the formation of single molecules in optical tweezer traps~\cite{liu2018building,zhang2020forming, cairncross2021assembly}. Additional experimental studies on thermal Na-Cs mixtures~\cite{Docenko2004} have led to the prediction of collisional properties and Feshbach resonances between Na and Cs~\cite{docenko2006coupling}. 

Here, we develop a cooling strategy for the simultaneous production of Bose-Einstein condensates of Na and Cs. We employ sympathetic cooling of Cs with Na in a two-stage process, using a magnetic quadrupole trap and a crossed optical dipole trap, which yields BECs with $8(1) \times 10^5$ Na and $3.5(1) \times 10^4$ Cs atoms (Fig.~\ref{fig:bec_formation}). Thanks to favorable trapping conditions and scattering properties that ensure good overlap between the species, our scheme does not require some of the complexities that have been common for the preparation of Cs condensates, such as degenerate Raman sideband cooling~\cite{vuletic1998degenerate} and magnetic levitation~\cite{weber2003bose}. In addition, we find that a Cs BEC overlapping with a Na BEC stays condensed for several seconds, constituting a stable quantum gas mixture that offers ideal conditions for the creation of novel many-body quantum systems.

A key motivation for the production of ultracold Na-Cs mixtures is the creation of ultracold ensembles of NaCs ground state molecules. NaCs has a permanent electric dipole moment of $d=4.6\,$Debye~\cite{aymar2005}, the largest of the chemically stable bialkali molecules~\cite{Zuchowski2010}. With NaCs, it will be possible to reach an effective range of dipole-dipole interactions, $a_\mathrm{d} = m d^2/(8 \pi \epsilon_0 \hbar^2)$~\cite{julienne2011universal}, of tens of micrometers, one order of magnitude larger than for NaK and two orders of magnitude larger than for KRb ground state molecules ($m$ denotes the molecular mass). NaCs is an exceptionally promising candidate for the realization of strongly correlated phases in dipolar quantum gases~\cite{lahaye2009physics, Baranov2012}, such as dipolar crystals~\cite{buchler2007strongly} and Mott insulators with fractional filling~\cite{Capogrosso2010}.

\section{Cooling strategy}\label{sec:sympathetic}


\begin{figure*} [t]
    \centering
    \includegraphics[width = 17.6 cm]{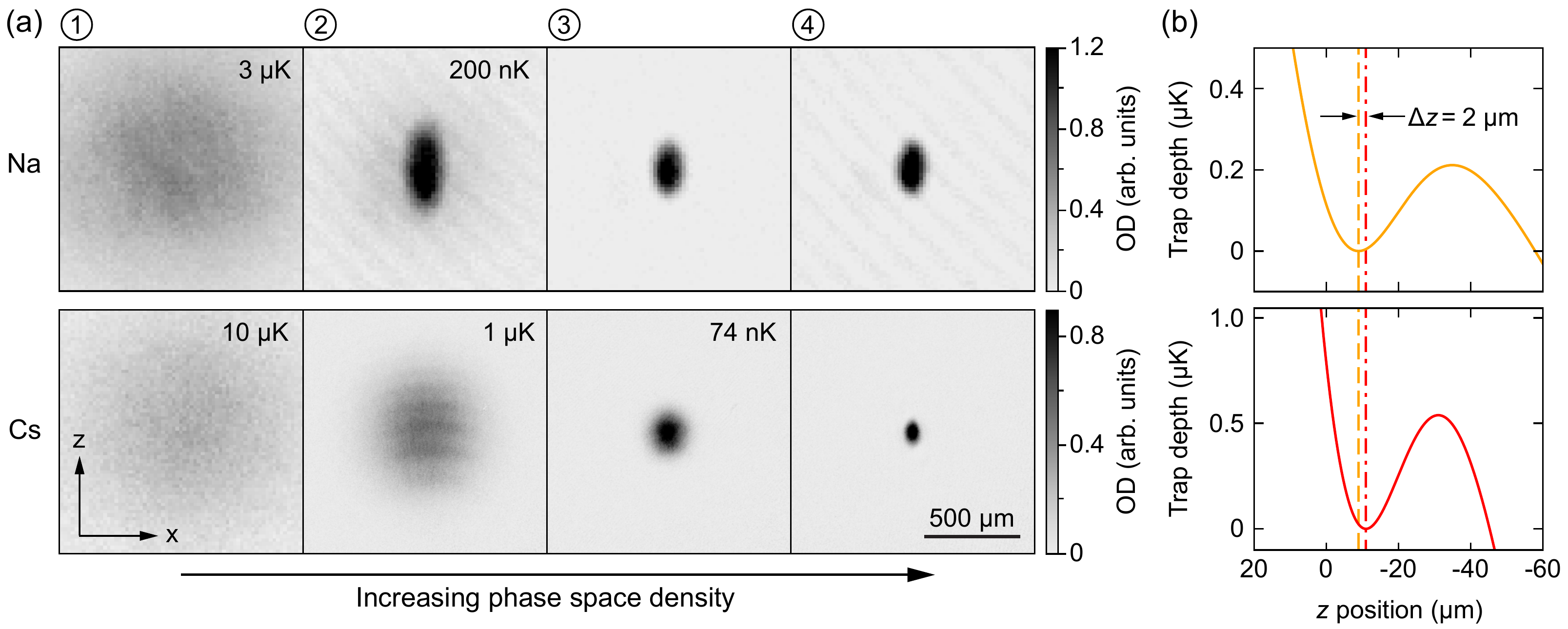}\\
    \caption{Na and Cs crossing the BEC phase transition. (a) Na (top) and Cs (bottom) clouds during  evaporative cooling. Phase-space densities increase from left to right. Each vertical pair of images corresponds to the same point in the evaporation sequence (see corresponding encircled numbers in Fig.~\ref{fig:timing}). Temperature labels in the upper right corner indicate the fitted temperature, in case a thermal component is present. All images are taken after $23.5\,$ms time-of-flight expansion. (b) Trapping potential for Na (top) and Cs (bottom) along the $z$-direction (gravity points in $-z$-direction) at the end of the evaporation sequence. The dashed (dot-dashed) line indicates the position of the trap minimum for Na (Cs).
    }
    \label{fig:bec_formation}
\end{figure*}

At first sight, the simultaneous preparation of ultracold gases of Na and Cs appears to be challenging due to seemingly conflicting requirements to achieve cooling towards quantum degeneracy. We demonstrate that the challenges can be overcome in an elegant way. 

For Cs, typically an all-optical cooling approach is chosen, in which the atomic ensemble is trapped in an optical dipole trap throughout the evaporative cooling sequence. Internal states of Cs that are magnetically trappable suffer from excessive two-body loss due to spin relaxation, which hindered early attempts to create BECs of Cs in magnetic traps \cite{Soding1998,Guery1998Euro,Thomas2003,Frye2019}. To avoid spin relaxation, Cs is prepared in its internal ground state, which is not magnetically trappable and necessitates optical trapping. However, in this state Cs ensembles are prone to significant three-body losses   \cite{Perrin1998,kerman2000beyond,Han2001}. To enable efficient evaporative cooling, specific magnetic fields close to one of the numerous Feshbach resonances are chosen to tune the scattering length to moderately positive values. Common working points are at $23\,$G~\cite{weber2003bose} and $894\,$G~\cite{berninger2013feshbach}, where the Cs intraspecies scattering lengths are $314\,a_0$ and $286\,a_0$, respectively. All-optical cooling has been the method of choice since the first demonstration of Bose-Einstein condensation of Cs \cite{weber2003bose}. 

For Na, cooling to ultracold temperatures typically relies on a magnetic trap. The high volume of magnetic traps allows for the evaporation of large atomic ensembles, and has been the well-established method to create Na BECs with millions of atoms since the first observation of Bose-Einstein condensation of Na~\cite{davis1995bose}. All-optical cooling of Na has been demonstrated~\cite{mimoun2010fast, jiang2013simple} but the resulting BECs are significantly smaller, and the optical dipole traps in the respective setups typically operate at trapping frequencies that would lead to unfavorably high densities and three-body loss rates for Cs.

Here, we demonstrate a scheme that ensures favorable parameters for both Na and Cs in a two-stage cooling process. Cs is sympathetically cooled by Na, first in an optically-plugged magnetic quadrupole trap~\cite{davis1995bose, naik2005optically, heo2011fast, park2012quantum}, and then in a crossed optical dipole trap. A similar approach has been previously employed to cool mixtures of Rb and Cs~\cite{mccarron2011dual}. In our setting, large Na clouds and favorable interspecies scattering properties ensure that Na serves as an efficient sympathetic coolant for Cs. 

In the first stage, the mixture is cooled in a magnetic quadrupole trap that is optically plugged with a $532\,$nm laser beam, which is blue-detuned for both Na and Cs to prevent Majorana losses at the center of the quadrupole field~\cite{davis1995bose}. The plug beam has a power of $10\,$W and a waist of $20\,\mu$m. For the second stage, both species are transferred to an optical dipole trap consisting of three beams in the horizontal plane (Fig.~2): two elliptical beams, labeled X ODT and Y ODT, at a wavelength of $1064\,$nm, and a round beam, labeled R ODT, at a wavelength of $1070\,$nm that passes through the crossing point of the X and Y ODT (beam sizes are provided in Table~\ref{tab:dipole_traps} in Appendix~\ref{sec:appendix_model}). R ODT acts as a reservoir beam to enhance the trap volume, ensuring that around 30\%
of the Na atoms are transferred from the magnetic trap to the optical trap.

The wavelengths of the dipole trap beams are chosen such that the trap frequencies for Na and Cs are closely matched. The differences in mass and optical polarizability almost perfectly compensate each other, ensuring a high degree of overlap of the mixture. At the end of the evaporation sequence, the trap frequencies for Na are $\{\omega^\mathrm{Na}_x, \omega^\mathrm{Na}_y, \omega^\mathrm{Na}_z\} = 2\pi \times \{37(1), 28(1), 147(6) \}\,$Hz and the trap frequencies for Cs are $\{\omega^\mathrm{Cs}_x, \omega^\mathrm{Cs}_y, \omega^\mathrm{Cs}_z\} = 2\pi \times \{33(1), 26(1), 123(5)\}\,$Hz. This leads to a differential gravitational sag between the equilibrium positions of the two clouds that is as small as $\Delta z = 2\,\mu$m (see Fig.~\ref{fig:bec_formation} (b)). Additional details and numerical modeling of the dipole trap are discussed in Appendix~\ref{sec:appendix_model}.


\begin{figure} [t]
    \centering
    \includegraphics[width = 8.6 cm]{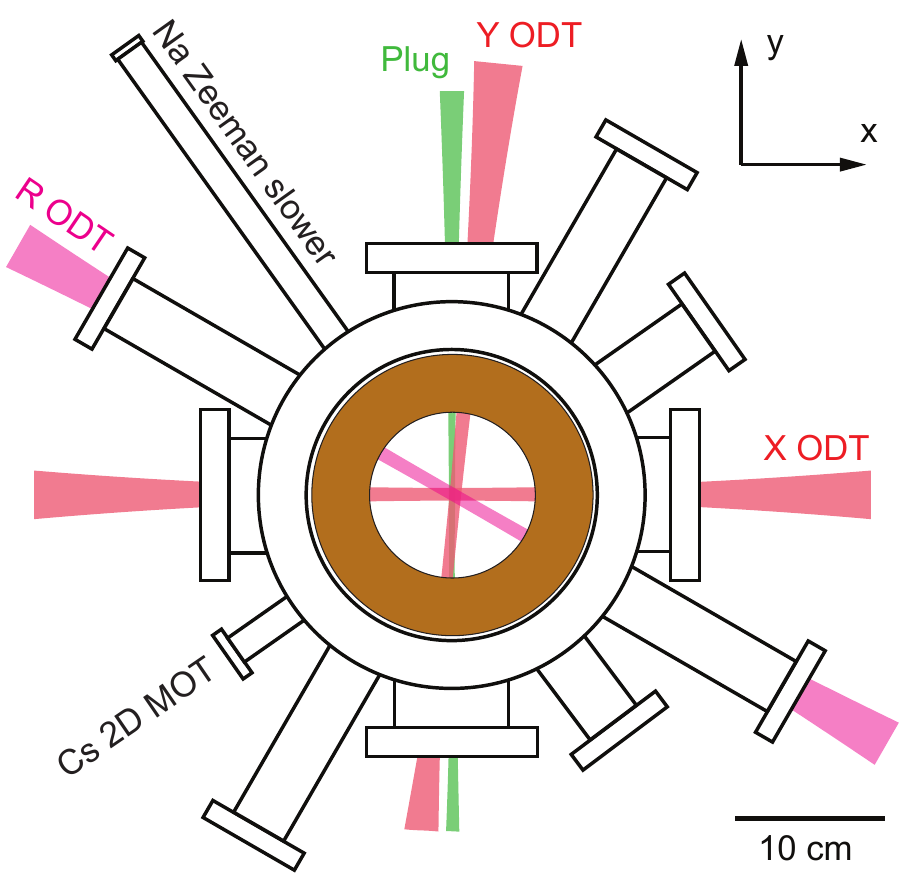}\\
    \caption{Schematic of the experiment chamber, as viewed from top. Cold beams of Na (Cs) atoms enter the chamber from a Zeeman slower (2D MOT). Additional laser beams for absorption imaging (not shown) enter the chamber on the $x$-axis and the $y$-axis. The brown ring represents the magnetic field coils that can be switched between Helmholtz and anti-Helmholtz configurations. 
    }
    \label{fig:coilandodt}
\end{figure}


\begin{figure}[t]
    \centering
    \includegraphics[width = 8.6 cm]{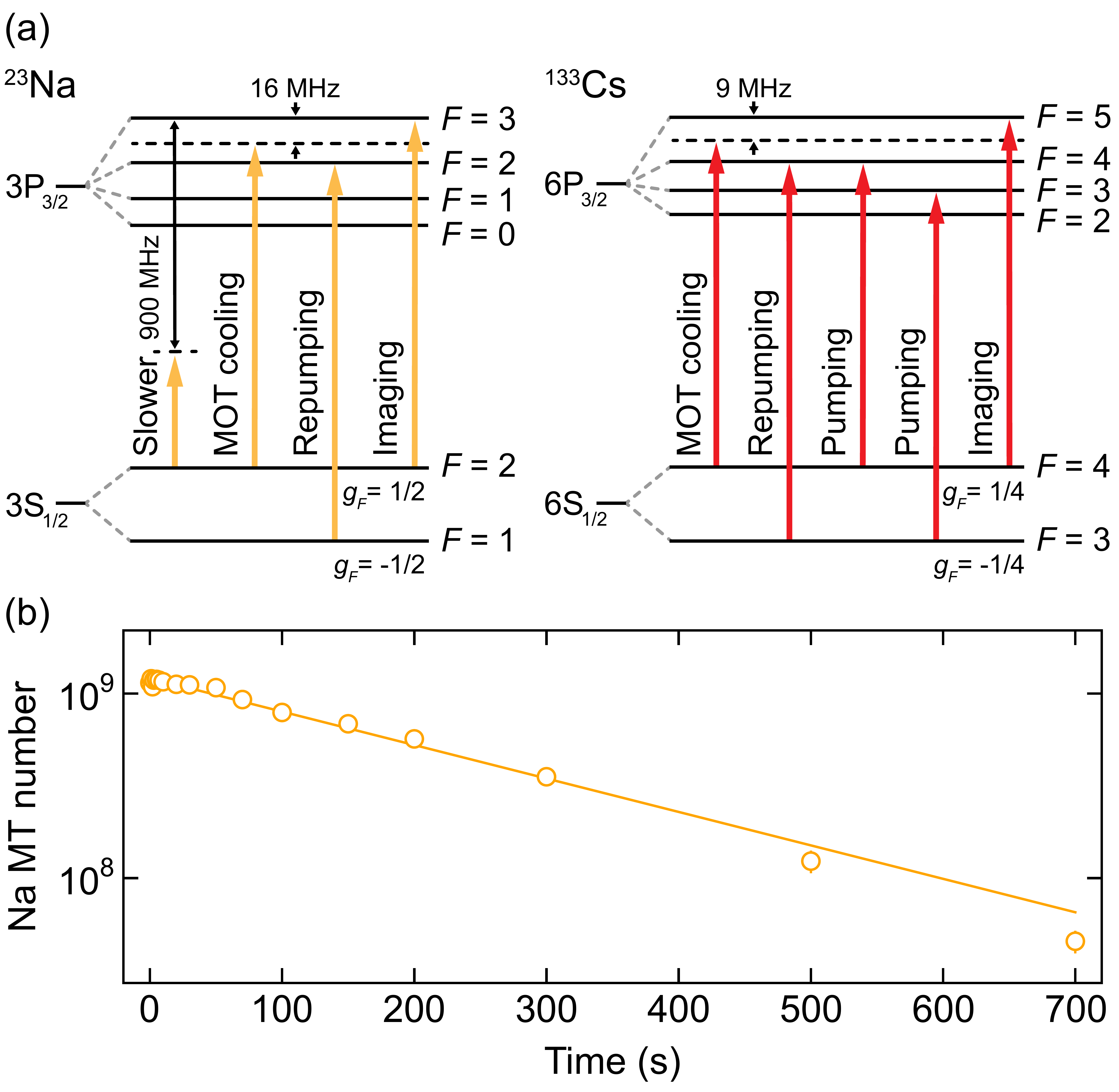}\\
    \caption{Preparation of Na-Cs mixtures. (a) Level diagrams for Na and Cs indicating the transitions used for laser cooling, repumping, and imaging. (b) Lifetime of Na atoms in the magnetic trap. 
    }
    \label{fig:levelsmotmt}
\end{figure}


\begin{figure*} [t]
    \centering
    \includegraphics[width = 17.6 cm]{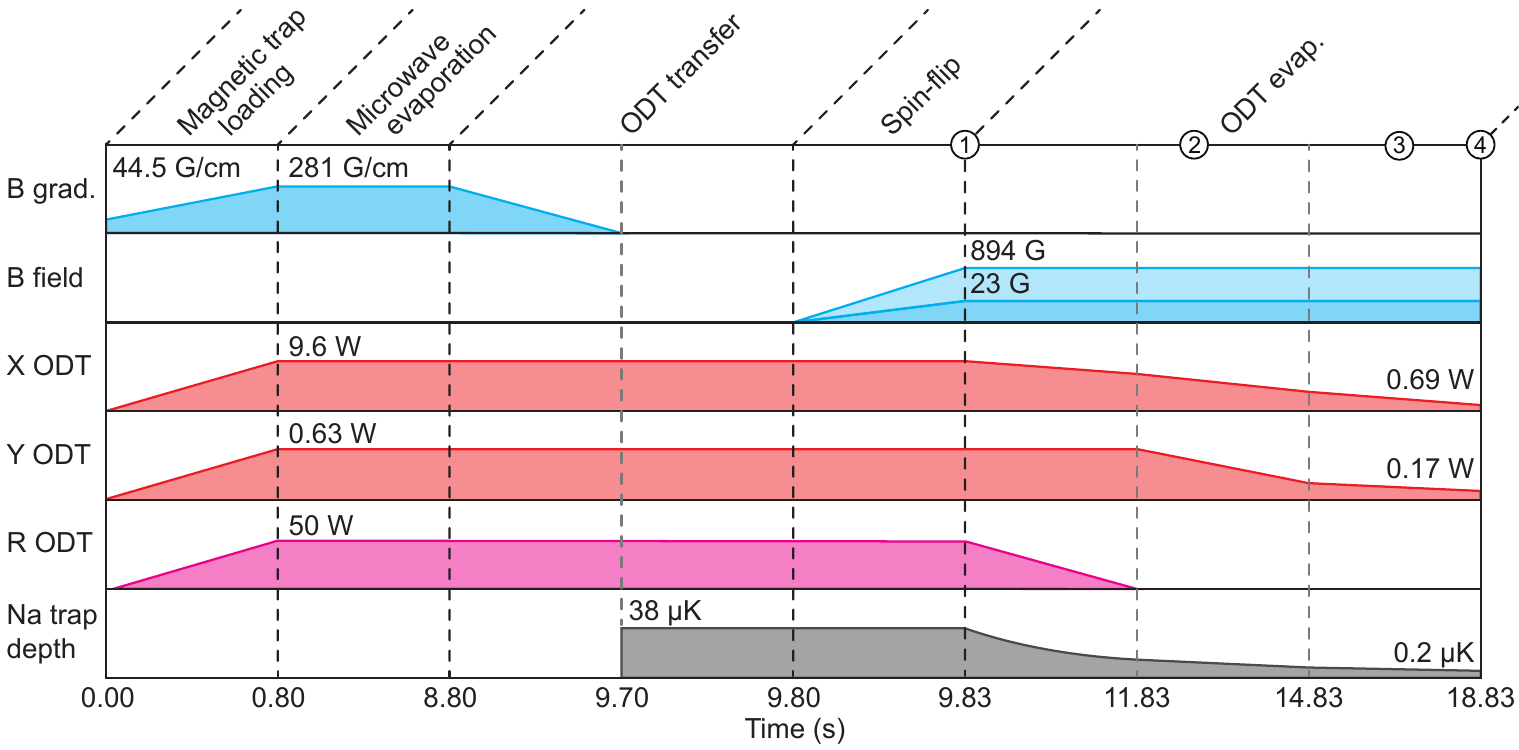}\\
    \caption{Timing diagram for simultaneous cooling of Na and Cs. Parameters are described in the main text. The last row indicates the trap depth for Na (taking into account the effect of gravity). The corresponding values for Cs are 164~$\mu$K at the beginning, and 0.5~$\mu$K at the end of the sequence. }
    \label{fig:timing}
\end{figure*}

\section{Sympathetic cooling}

Our experiment begins by loading about $3 \times 10^9$ Na atoms and $3 \times 10^7$ Cs atoms into a dual-species magneto-optical trap (MOT). The relevant energy levels and laser frequencies for laser cooling, repumping, and imaging of Na and Cs are shown in Fig.~\ref{fig:levelsmotmt} (a). For Na, we employ a dark-spot MOT~\cite{ketterle1993high} loaded from a spin-flip Zeeman slower~\cite{streed2006large}. For Cs, we use a bright MOT loaded from a 2D MOT~\cite{lam2020compact}. After loading the MOT for $4\,$s, an optical molasses is used to cool Na (Cs) to $36\,\mu$K ($6\,\mu$K), followed by optical pumping into the hyperfine ground state manifold $F=1$ ($F=3$). To enhance transfer to the magnetic trap, Cs is optically pumped into the magnetically trappable $| F , m_F \rangle = | 3, -3 \rangle$ state. Here, $F$ denotes the total angular momentum and $m_F$ its projection on the magnetic field axis. 

Na and Cs are simultaneously loaded into the optically-plugged magnetic quadrupole trap. Initially the trapping field is snapped on to a gradient of $44.5\,$G/cm, chosen to ensure the loading of spin-purified ensembles of Na (Cs) in the $| 1, -1 \rangle$ ($| 3, -3 \rangle$) state. We observe a magnetic trap lifetime of $240(8)\,$s (see Fig.~\ref{fig:levelsmotmt} (b)), limited by background pressure in the vacuum chamber. The experimental sequence that follows is illustrated in Fig.~\ref{fig:timing}. The trap is compressed to a gradient of $281\,$G/cm to increase the collision rate for efficient evaporative cooling. At the same time, all ODT beams are adiabatically increased to their maximum powers and remain on during evaporation in the magnetic trap~\cite{lin2009rapid}. During  compression, a microwave knife driving the Na $| 1, -1 \rangle \rightarrow | 2, -2 \rangle$ transition at $1680\,$MHz is kept on, setting a Na temperature cutoff of about $1.5\,$mK.  

\begin{figure} 
    \centering
    \includegraphics[width = 8.6 cm]{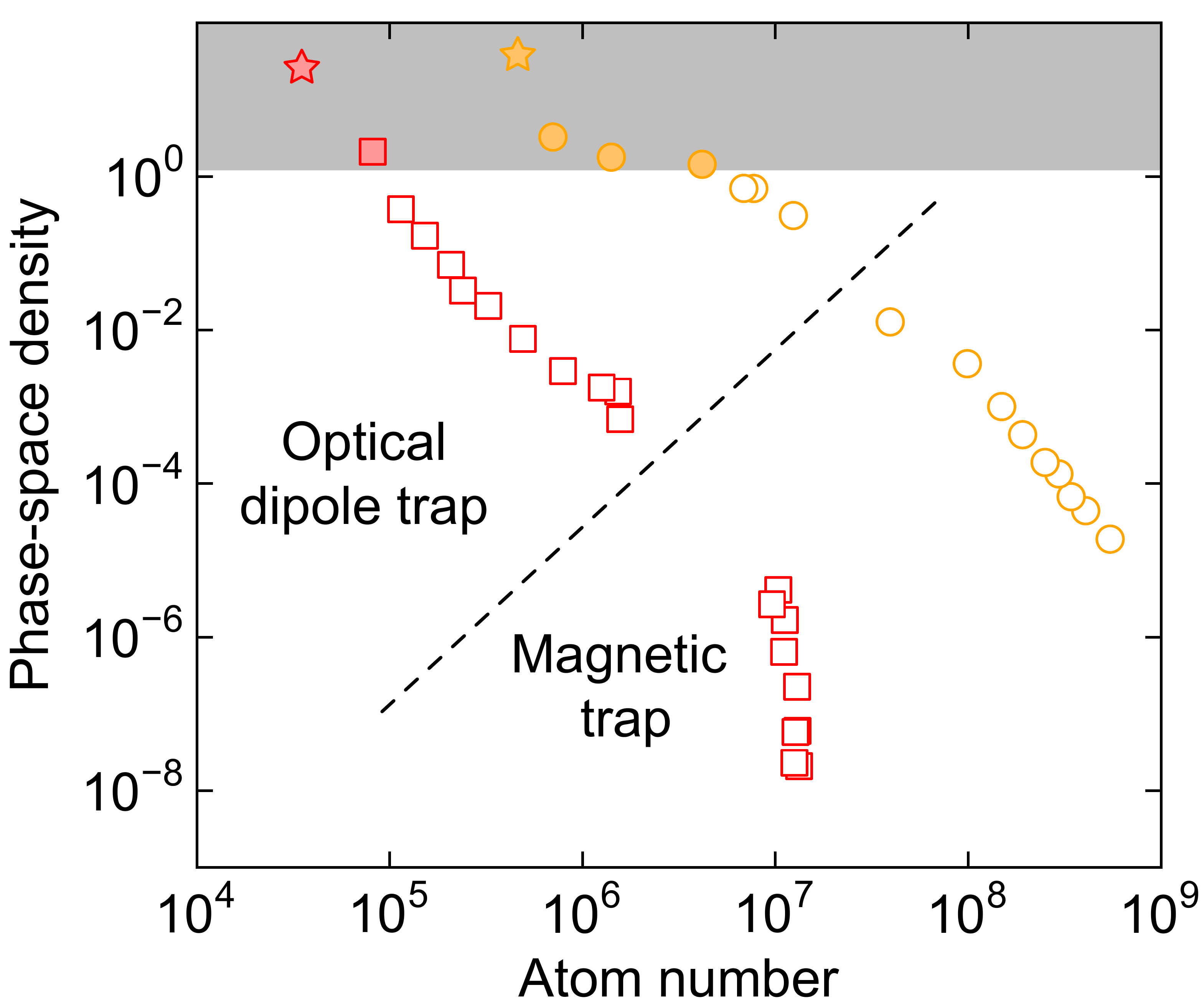}\\
    \caption{Phase-space density versus atom number for Na (orange circles) and Cs (red squares) during the cooling process, measured for the working point at $23\,$G. Empty points indicate a thermal gas, filled points indicate that a condensate is present, and stars indicate the final atom number of the BEC for each species. Error bars are smaller than the size of the data points. The grey shaded region indicates phase-space densities for which condensation is expected.
    }
    \label{fig:psdvsn}
\end{figure}

Forced microwave evaporation is applied to Na on the $| 1, -1 \rangle \rightarrow | 2, -2 \rangle$ transition, while Cs is sympathetically cooled. Fig.~\ref{fig:psdvsn} shows the evolution of phase-space density (PSD) for Na and Cs in the course of the cooling sequence. Na starts at a PSD of $2(1) \times 10^{-5}$ with $5.4(2) \times 10^8$ atoms at $180(10)\,\mu$K and Cs starts at a PSD of $2.1(3) \times 10^{-8}$ with $1.3(1) \times 10^7$ atoms at $250(8)\,\mu$K. The frequency of the microwave knife is linearly increased to $1760\,$MHz within $8\,$s. For Na, this results in a PSD of 0.013(1) with $3.9(1) \times 10^7$ atoms at $22.6(2)\,\mu$K, corresponding to an evaporation efficiency $\alpha \approx 2.5$~\footnote{ The evaporation efficiency is defined as $\alpha = - \ln( \mathrm{PSD}_{\mathrm{final}} / \mathrm{PSD}_{\mathrm{initial}} )/ \ln( N_{\mathrm{final}} / N_{\mathrm{initial}} )$}.  For Cs, the PSD reaches $3(1) \times 10^{-6}$ with $9.6(3) \times 10^6$ atoms at $79(6)\,\mu$K, retaining almost all Cs atoms. The cooling efficiency for Cs is $\alpha \approx 16$, indicating highly efficient sympathetic cooling with minimal atom loss. Calculating the cloud sizes in our magnetic trap, we find that the Cs cloud is entirely enclosed by the Na cloud, which ensures good thermal contact and efficient sympathetic cooling of Cs by Na. However, the thermalization of Cs lags behind Na due to a relatively small interspecies scattering length, as discussed below. At the end of evaporative cooling in the magnetic trap, the Cs temperature is three to four times higher than the Na temperature. This leads to a reduction of peak density, which has the positive side effect that two- and three-body losses of Cs are reduced.

The mixture is transferred into a pure optical dipole trap by ramping the magnetic quadrupole field to zero and switching off the optical plug beam. A homogeneous magnetic field is ramped to either $23\,$G or $894\,$G. At the same time, a $5\,$MHz radio frequency (RF) field induces a Landau-Zener sweep, transferring the Na and Cs atoms into their internal ground states $| 1, 1\rangle$ and $| 3,3 \rangle$, respectively. Typically, $1.2(1)\times 10^7$ Na atoms at $3.5(2)\,\mu$K and $1.6(3)\times 10^6$ Cs atoms at $10(1)\,\mu$K are transferred to the dipole trap. Losses of Na during the transfer are mostly the result of free evaporation; losses of Cs are dominantly caused by two- and three-body collisions prior to the Landau-Zener sweep and before the working points at $23\,$G or $894\,$G are reached. 

\section{Simultaneous Condensation}\label{sec:condensates}


In the optical dipole trap, we continue evaporative cooling to achieve condensation of Na and Cs. Fig.~\ref{fig:bec_formation} (a) shows the cooling progression towards BEC for both species. Evaporation of Na is achieved in three steps that reduce the dipole trap depth in piecewise linear ramps (Fig.~\ref{fig:timing}), while Cs is sympathetically cooled by Na. Direct evaporation of Cs is strongly suppressed as the trap depth is three to four times higher than for Na. In the first step, R ODT is ramped to zero, and X and Y ODT are tuned to equal trap depths. Then, two additional linear ramps reduce X and Y ODT to their final trap depths.

Na condenses at the end of the first ramp at a critical temperature of about $500\,$nK. The critical temperature for Na condensation is significantly higher than for Cs due to the larger atom number. At the beginning of evaporation in the ODT, Na has a PSD of 0.31(4). At the end of the evaporation sequence, we obtain Na BECs with $8(1) \times 10^5$ atoms. Cs condenses at the end of the third step at a critical temperature of about $50\,$nK. At the beginning of evaporation in the ODT, Cs starts with a PSD of $6.9(4) \times 10^{-4}$. At the end of the sequence, we obtain a quasi-pure Cs condensate with $3.5(1) \times 10^4$ atoms. The evaporation efficiency is about $\alpha \approx 2.7$, limited by three-body losses.

We calculate Thomas-Fermi radii in the $z$-direction for the Na and Cs BECs of $5.1(2)$ and $2.7(1)\,\mu$m, respectively. Given the  gravitational sag of $2\,\mu$m, we expect that the Na BEC encloses the Cs BEC. This is supported by  cross-thermalization and lifetime measurements described below, which indicate that Na and Cs BECs are miscible and overlapping. In addition, we have observed the formation of Feshbach molecules from overlapping condensates of Na and Cs, which will be discussed in subsequent work~\cite{lam2021feshbach}. 

\section{Cross Thermalization}\label{sec:thermalization}

\begin{table}
    \centering
    \caption{Intraspecies $s$-wave scattering lengths for Na $|1, 1\rangle$ and Cs $|3, 3\rangle$ and measured interspecies scattering lengths. }
    \label{tab:scattering_lengths}
    \begin{tabular}{c c c c} \hline \hline
    Scattering length ($a_0$)& $23.0\,$G & $894.0\,$G & Ref.\\ \hline
    $a_{\mr{Na}}$ & 54.5 & 64.3 & \cite{knoop2011feshbach}\\
    $a_{\mr{Cs}}$ & 314 & 286 & \cite{berninger2013feshbach} \\
    $a_{\mr{NaCs}}$ & 18(4) & 29(4)& This work\\ \hline \hline
    \end{tabular}
\end{table}


\begin{figure}
    \centering
    \includegraphics[width = 8.6 cm]{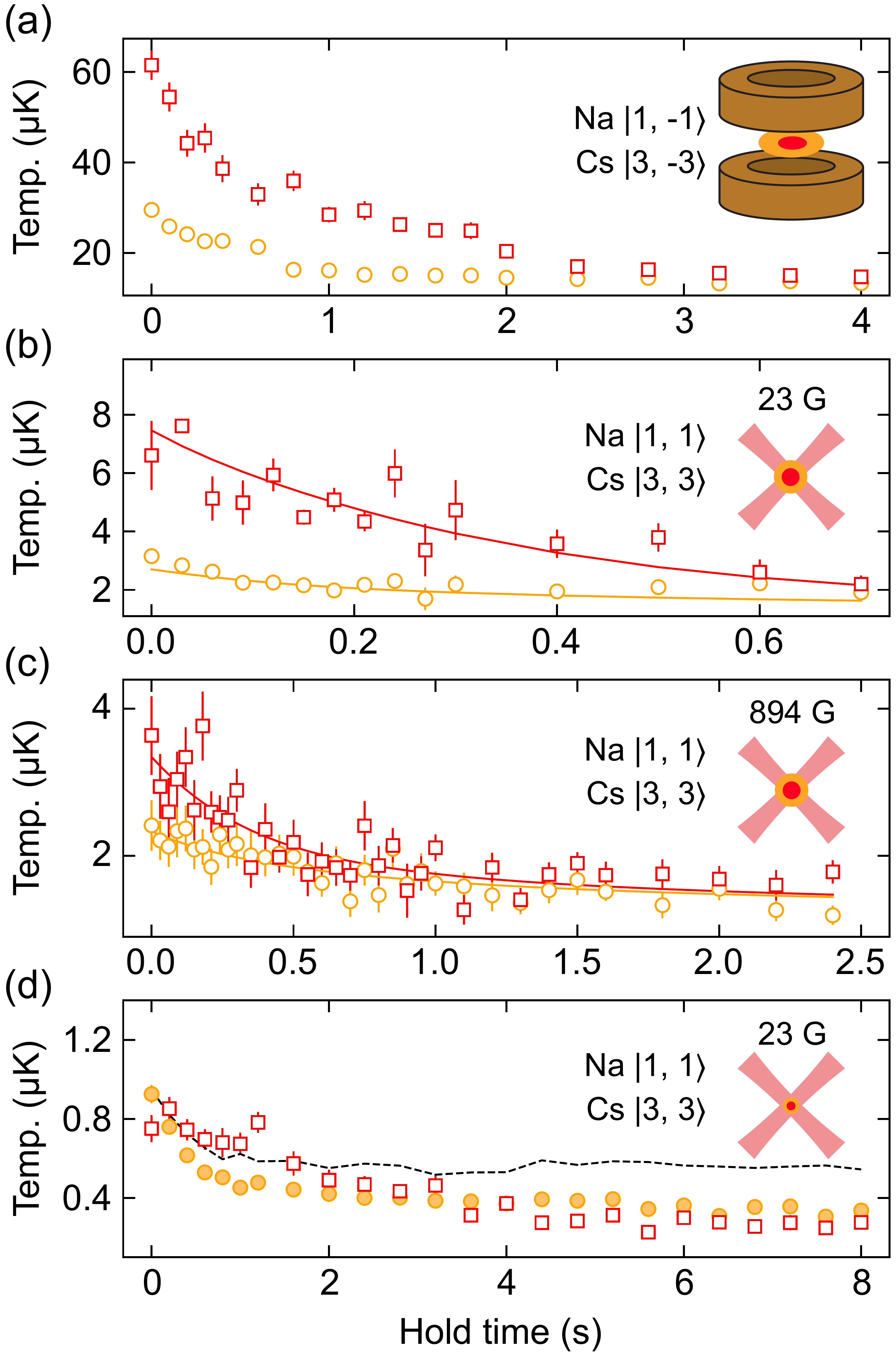}\\
    \caption{Thermalization of Na and Cs.  Na (Cs) data is shown as orange circles (red squares). Filled (empty) data points indicate the presence of a condensate (thermal cloud), error bars indicate the standard error of the mean of three experimental runs. Solid lines show the fit of the classical kinetic model. (a) Thermalization in the magnetic trap.  (b)  Thermalization between Na and Cs thermal clouds in the ODT at $23\,$G. (c) Thermalization between Na and Cs thermal clouds in the ODT at 894~G. (d) Thermalization between a Na BEC and a Cs thermal cloud. The dashed line indicates the critical temperature for a Na BEC, calculated from the atom numbers and trap parameters. During hold time, the Cs thermal cloud cools to about half of the Na critical temperature.}
    \label{fig:thermalization}
\end{figure}


\begin{figure*}
    \centering
    \includegraphics[width = 17.6 cm]{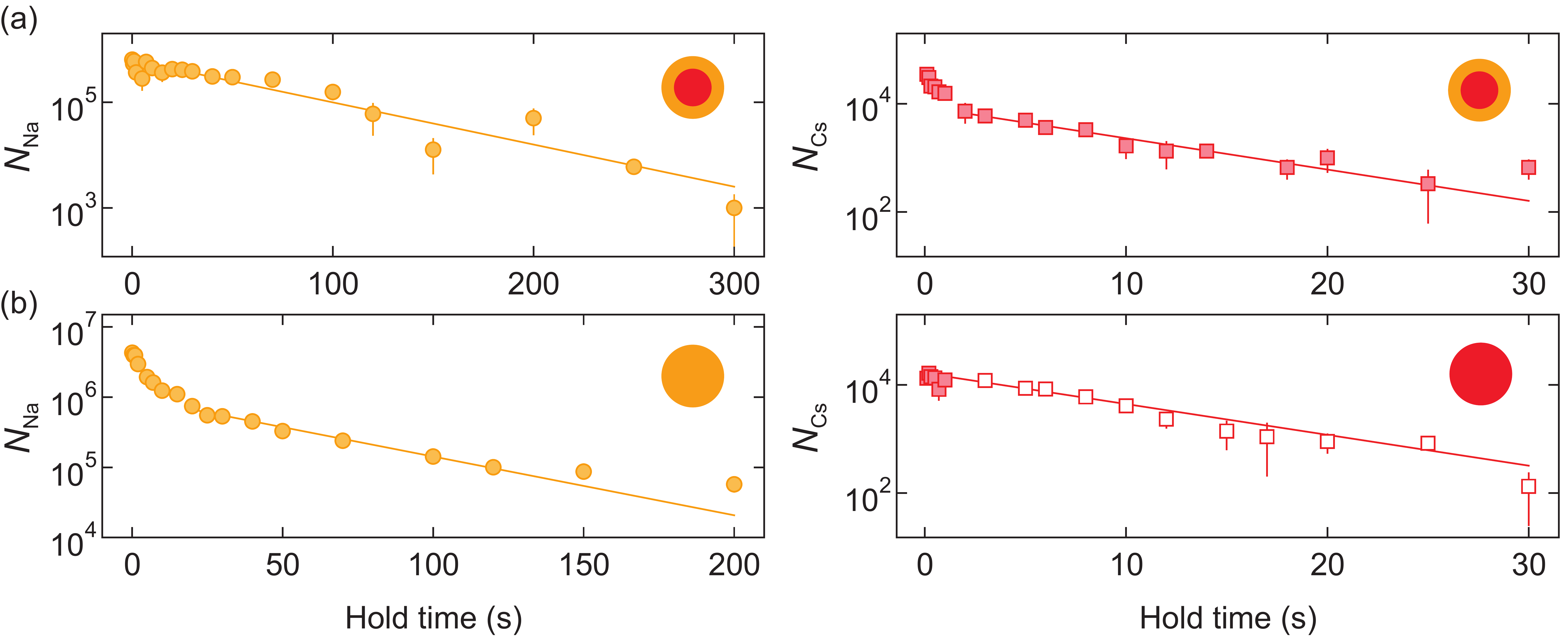}\\
    \caption{Lifetimes of Na and Cs BECs at $23\,$G. (a) Decay of  overlapping Na and Cs condensates. Orange circles (red squares) represent Na (Cs). (b) Decay of standalone Na and Cs condensates.  Filled (empty) data points indicate the presence of a BEC (thermal cloud). Dashed lines represent fits that take into account one- and three-body decay, solid lines represent pure one-body decay. }
    \label{fig:bec_lifetimes_fw}
\end{figure*}

To investigate the scattering properties of the mixture, we study cross thermalization between Na and Cs both in the magnetic quadrupole trap and the ODT. We initiate the thermalization dynamics by suddenly reducing the trap depth for Na and record temperatures of both clouds as a function of hold time, as shown in Fig.~\ref{fig:thermalization}. To analyze the data, we employ a classical kinetic model~\cite{olson2013optimizing}. With the model, we fit the interspecies scattering cross section for Na and Cs, which allows us to extract the magnitude of the $s$-wave scattering length (see Appendix~\ref{sec:appendix_kineticmodel}).

In Fig.~\ref{fig:thermalization} (a), cross-thermalization is shown for thermal Na-Cs mixtures in the magnetic trap. For this data, different from the standard sequence, we load a small cloud of about $2 \times 10^6$ Cs atoms into the magnetic trap to reduce the heat load from Cs on Na. Thermalization dynamics are started by applying a fast microwave ramp that abruptly removes a portion of hot Na atoms, and we record the temperature of the mixture for various hold times. The data shows that Na and Cs thermalize on a time scale of about $2\,$s. This is sufficiently fast to enable sympathetic cooling in the magnetic trap within the $8\,$s time scale of the microwave evaporation stage. We limit our analysis to this qualitative observation, as Cs samples a range of magnetic fields in the trap, which comes with a significant variation of the intraspecies scattering length~\cite{Frye2019}. The following data on cross thermalization in the ODT allows us to extract quantitative information.

In the ODT, we first study thermalization of thermal mixtures of Na and Cs for magnetic fields of $23\,$G and $894\,$G, shown in Figs.~\ref{fig:thermalization} (b) and (c), respectively. The dynamics are initiated by turning off the R ODT within $100\,$ms before the hold time starts, and we subsequently observe the evolution of temperatures. We fit the data using the classical kinetic model, which yields an interspecies scattering length of $a_{\mathrm{NaCs}}=18(4)\,a_{0}$ at $23\,$G and $a_{\mathrm{NaCs}}=29(4)\,a_{0}$ at $894\,$G~\footnote{While the fitting model is not sensitive to the sign of the scattering length, we quote the positive sign here in accordance with Refs.~\cite{docenko2006coupling, hood2020multichannel, Tiemann2021}}. While cross thermalization is slow, the Na-Cs mixture is miscible and interspecies three-body losses are strongly suppressed (see below) thanks to the relatively small values of the interspecies scattering length. The condition for miscibility of Na and Cs BECs is given by $\sqrt{g_{\mr{Na}}g_{\mr{Cs}}}/g_{\mr{NaCs}} > 1$  \cite{riboli2002topology} \footnote{The interspecies coupling constant is given by $g_{\mr{NaCs}} = 2 \pi \hbar^2 a_{\mr{NaCs}} \left(m_{\mr{Na}} + m_{\mr{Cs}} \right)/m_{\mr{Na}}m_{\mr{Cs}}$ and the intraspecies coupling constants are given by $g_{\mr{Na}} = 4 \pi \hbar^2 a_{\mr{Na}}/m_{\mr{Na}}$ and $g_{\mr{Cs}} = 4 \pi \hbar^2 a_{\mr{Cs}}/m_{\mr{Cs}}$}. Using the known scattering lengths for Na and Cs and the data at $23\,$G and $894\,$G measured here (see Table \ref{tab:scattering_lengths}), we find this ratio to be approximately $5$ and $3.3$, respectively. As a result, overlapping BECs of Na and Cs are deeply in the miscible regime. 

Finally, we study thermalization in the ODT between a Na BEC and a thermal cloud of Cs at $23\,$G, shown in Fig.~\ref{fig:thermalization} (d). Again, the dynamics are started by turning off the R ODT within $100\,$ms. For the Na BEC, we determine the temperature from the observed condensate fraction, which increases as a function of hold time, indicating cooling of Na due to free evaporation. Similarly, we find that Cs becomes colder. We note that free evaporation of Cs is strongly suppressed, as the truncation parameter for Cs is approximately 30 at this point~\cite{ketterle1996evaporative}. Consequently, the reduction of the Cs temperature is likely the result of sympathetic cooling of the thermal Cs cloud by the Na BEC. Calculating the cloud radii along the $z$-direction, we obtain for the thermal Cs cloud a radius of $2.2(1) \, \mu$m and for the Na BEC a Thomas-Fermi radius of $5.1(2)\, \mu $m. For the trap parameters at this point in the experimental sequence, the differential gravitational sag is less than  $1 \, \mu$m. The Cs thermal cloud is fully enclosed by the Na BEC, and thermal contact between the miscible and overlapping clouds mediates the cooling.

\section{Lifetimes}

To characterize the stability of the mixture, we investigate the lifetimes of overlapping Na and Cs condensates (Fig.~\ref{fig:bec_lifetimes_fw}). For comparison, we also measure the lifetimes of standalone condensates of both species. The measurements are performed in the crossed X and Y ODT at a magnetic field of $23\,$G. Remarkably, we find that a Cs BEC that overlaps with a Na BEC lives significantly longer than a standalone Cs BEC under otherwise identical trapping conditions.

For the case of overlapping BECs, we detect condensed clouds of Na and Cs throughout the entire range of hold times, $300\,$s and $30\,$s, respectively, as shown in Fig.~\ref{fig:bec_lifetimes_fw} (a). Longer hold times for Cs BECs are not recorded as the signal-to-noise limit of our Cs imaging system was reached. At the given densities, the decay of Na is dominated by one-body loss with a 1/$e$-lifetime of $54(3)\,$s, which is shorter than the lifetime in the magnetic trap due to technical noise in the ODT. The decay of Cs is also dominated by one-body losses, with the faster initial decay caused by continued free evaporation after the ODTs are ramped to their final values. We find a Cs $1/e$-lifetime of $7.5(6)\,$s after the faster initial decay. We see no indication of significant interspecies three-body processes of the types Na-Na-Cs and Na-Cs-Cs.


For the stand-alone BECs, Na remains condensed for the entire observation time, while the Cs BEC turns thermal after $2\,$s of hold time, as shown in Fig.~\ref{fig:bec_lifetimes_fw} (b). To form a standalone Na BEC, we only load a Na MOT and proceed with evaporative cooling in a sequence similar to that described above, producing condensates of $5.1(1) \times 10^6$ atoms after R ODT is ramped off. We observe initial fast losses, which are followed by one-body decay with a $1/e$-lifetime of $52(5)\,$s. To form a stand-alone Cs BEC, we apply a magnetic field gradient in the final stages of ODT evaporation that lowers the trap depth for Na, removes Na, and leaves behind a trapped Cs BEC. We observe dominant one-body decay. Likely, fast initial losses due to free evaporation occur while Na is removed from the trap which is not recorded here. We find a Cs $1/e$-lifetime of $7.6(4)\,$s. 

The presence of a Na BEC significantly extends the lifetime of the Cs BEC, at least by a factor of ten. We attribute the extended lifetime of Cs BECs to continued sympathetic cooling, which counteracts heating processes in the ODT, indicating good thermal contact and overlap between Na and Cs BECs. 


%

\section{Conclusion}

We have created overlapping BECs of Na and Cs. We have developed a cooling strategy that relies on sympathetic cooling of Cs, using Na as a coolant in a two-stage process that involves magnetic and optical trapping. The interactions between Na and Cs are found to be moderately repulsive, which leads to miscible bosonic quantum gas mixtures that remain stable and condensed for tens of seconds. 
Ultracold mixtures of Na and Cs offer exciting prospects for many-body physics with bosonic quantum gas mixtures, including  studies of overlapping superfluids~\cite{ho1996binary, pu1998properties, sinatra1999dynamics, papp2008tunable, pattinson2013equilibrium, lee2016phase} and polaron physics \cite{hohmann2015neutral,grusdt2017strong} using Cs as heavy impurities immersed in a Na BEC. Overlapping Na and Cs BECs offer ideal starting conditions for the creation of ultracold gases of NaCs molecules via association from overlapping bulk gases, as well as from dual bosonic Mott insulators in an optical lattice \cite{damski2003creation}. 

\section*{Acknowledgements}

We thank R.~Grimm and E.~Tiemann for fruitful discussions and S.~Cornish, M.~Kwon, and T.~Yefsah for helpful comments on the manuscript. We also thank David Grabovsky for experimental assistance in the early stages of the experiment. This work was supported by an NSF CAREER Award (Award No.~1848466) and a Lenfest Junior Faculty Development Grant from Columbia University. S.W.~acknowledges additional support from the Alfred P.~Sloan Foundation. I.S.~was supported by the Ernest Kempton Adams Fund. C.W.~acknowledges support from the Natural Sciences and Engineering Research Council of Canada (NSERC) and the Chien-Shiung Wu Family Foundation. 

\appendix

\section{Dipole Trap Model} \label{sec:appendix_model}

To describe our optical dipole trap potential, we employ a model calibrated by measuring trap frequencies with the Na BEC. For the optical trapping potential of a laser beam we assume (including the effects of both D1 and D2 lines and omitting the rotating wave approximation) \cite{grimm2000optical}, 
\begin{equation}
\begin{split}
U(\mathbf{r})  =& -\frac{ \pi c^2}{ \omega_{\mr{D2}}^3} \left( \frac{\Gamma_{\mathrm{D2}}}{\omega_{\mr{D2}} - \omega_{\mr{l}}} + \frac{\Gamma_{\mr{D2}}}{\omega_{\mr{D2}} + \omega_{\mr{l}}} \right) I(\mathbf{r}) \\ 
& - \frac{\pi c^2}{2 \omega_{\mr{D1}}^3} \left( \frac{\Gamma_{\mr{D1}}}{\omega_{\mr{D1}} - \omega_{\mr{l}}} + \frac{\Gamma_{\mr{D1}}}{\omega_{\mr{D1}} + \omega_{\mr{l}}} \right) I(\mathbf{r}),
\label{dipole_potential}
\end{split}
\end{equation}
where $\omega_{\mr{l}}$ is the trapping laser frequency, $I(\mathbf{r})$ the Gaussian intensity profile of the laser beam, $\omega_{\mr{D2}}$ ($\omega_{\mr{D1}}$) is the transition frequency, and $\Gamma_{\mr{D2}}$ ($\Gamma_{\mr{D1}}$) is the linewidth of the atomic D2 (D1) line (see~\cite{steck2000sodium, steck2003cesium}).  We model the intensity of individual laser beams as
\begin{equation}
I(\mathbf{r}) = \frac{2 P_0}{\pi w_{\mr{h}}w_{\mr{v}}} e^{
-2 \left(x \sin\theta/w_{\mr{h}}\right)^2 
-2 \left(y \cos\theta/w_{\mr{h}}\right)^2 
-2 \left(z/w_{\mr{v}}\right)^2 
},
\end{equation}
where $P_0$ is the measured power of the beam, $w_{\mr{h}}$ ($w_{\mr{v}}$) is the horizontal (vertical) beam waist, and $\theta$ is the angle the beam makes with the $y$-axis.  The beam waists and maximum powers of the different dipole traps are listed in Table~\ref{tab:dipole_traps}.  To calculate the trap frequencies, we combine the trapping effects of each beam. The resulting trap frequencies are given by
\begin{equation}
\begin{split}
\omega_x & = \sqrt{ \sum_i \frac{4 \sin^2\theta_i }{m w^2_{\mr{h}_i}} U_i(0, 0, z_{\mr{min}}) }, \\
\omega_y & = \sqrt{ \sum_i \frac{4\cos^2\theta_i }{m w^2_{\mr{h}_i}} U_i(0, 0, z_{\mr{min}})  }, \\
\omega_z & = \sqrt{ \sum_i \left(1 - 4 \left(\frac{z_{\mr{min}}}{w_{\mr{v}_i}}\right)^2\right) \frac{4 U_i(0, 0, z_{\mr{min}})}{m w^2_{\mr{v}_i}} },
\end{split}
\end{equation}
where the sum runs over the trapping beams, $m$ is the mass of the atom, and $z_{\mr{min}}$ is the $z$-position of the potential minimum, found numerically, accounting for gravity.  The  $1 - 4 \left(z_{\mr{min}}/w_{\mr{v}_i}\right)^2$ term accounts for how the potential is distorted by gravity reducing the trap frequency~\cite{hung2011situ}.  Importantly, these terms capture how gravity breaks the symmetry of the trap frequencies.  Without gravity, $\omega_{\mr{Cs}} = 0.93 \omega_{\mr{Na}}$ (based on Eq.~\eqref{dipole_potential}), but due to their larger mass, the Cs atoms sit lower in the optical potential than the Na atoms and experience a lower light intensity than the Na atoms.  Although the relative gravitational sag given by the model is small, $2 \, \mu$m at the end of evaporation, the absolute gravitational sag is significant: $9 \,\mu$m ($11 \,\mu$m) for Na (Cs), compared to a beam waist of 36$\, \mu$m.  As a result, Cs atoms feel 7\% lower optical intensity from the $y$-beam than the Na atoms, further reducing the Cs $z$ trap frequency.  Since the atoms sit about a third of a beam waist below the beam center, the gravitational sag does not follow the usual $\Delta z = g / \omega_z^2$ formula as the underlying harmonic approximation is not valid.

\begin{table}
     \caption{Dipole trap parameters. The X and Y ODT beams are generated from a narrow-line single-mode Nd:YAG laser at $1064\,$ nm (Coherent Mephisto MOPA) and the R ODT beam from a multi-mode Yb fiber laser at $1070\,$ nm (IPG YLR-200-LP-AC). Beam waists are obtained from a fit to trap frequencies measured with a Na BEC at different points in the evaporation sequence. }
    \centering
    \begin{tabular}{cccc} \hline \hline
    {} &  \multicolumn{2}{c}{Waist ($\mu$m)} & {}\\
        Beam & Horizontal & Vertical  & Peak power (W) \\ \hline
        X ODT & 213(3) & 105(10) & 9.6 \\
        Y ODT & 155(1) & 37(1) & 0.63\\
        R ODT & 260(30) & 260(30) & 50 \\ \hline \hline
    \end{tabular}
    \label{tab:dipole_traps}
\end{table}

\begin{table*} [t]
    \centering
    \caption{Measured and modeled trap frequencies for the Na BEC. The fits are only performed for the X and Y ODT beams as R ODT is switched off early on in the evaporation sequence.}
    \begin{tabular}{*8c} 
    \hline \hline
    {} & {} & \multicolumn{2}{c}{$\omega_x / 2 \pi \,$(Hz)} &  \multicolumn{2}{c}{$\omega_y / 2 \pi \,$(Hz)} & \multicolumn{2}{c}{$\omega_z / 2 \pi \,$(Hz)}\\
     P$_x$ (W) & P$_y$ (W) & Expt. & Model & Expt. & Model & Expt. & Model \\ \hline 
    4.04 & 1.65 & 123 (1) & 121 & 70 (1) & 69.5 & 538 (4) & 530 \\
    1.97 & 0.838 & 85.5 (0.6) & 86.0 & 49.5 (0.7) & 48.5 & 363 (7) & 375 \\
    1.38 & 0.656 & 75.5 (0.4) & 76.0 & 43 (1) & 40.7 & 307 (7) & 330 \\
    0.649 & 0.297 & 50.5 (0.3) & 50.3 & 24.8 (0.6) & 27.8 & 211 (4) & 212 \\
    \hline \hline
    \end{tabular}
    \label{tab:waist_fitting}
\end{table*}

To measure the $x$ and $y$ trap frequencies in the experiment, we use a magnetic field gradient to offset the Na BEC from the trap minimum.  The gradient is suddenly turned off, inducing oscillation of the atoms along one coordinate axis. To measure the $z$ trap frequency, we use a brief free-fall to offset the Na BEC from the trap minimum. The dipole trap is suddenly turned back on, inducing oscillations in the $z$ direction. To obtain the dipole trap power inside the vacuum chamber, we measure the laser power on either side of the chamber and use the geometric mean of the two measurements, accounting for losses due to viewport reflectivity and absorption. The laser powers, along with measured Na trap frequencies, are listed in Table~\ref{tab:waist_fitting}. 

With the parameters determined using the Na BEC, the model is applied to Cs to determine the respective trap frequencies and depths. The model accurately predicts the trap bottom for Cs, as a cross-check of its accuracy.

\section{Kinetic Model} \label{sec:appendix_kineticmodel}
The evolution of the atom number ($N$) and the energy ($E$) of Na and Cs during  thermalization in the optical dipole trap is modelled by the following differential equations (superscript for the atomic species is omitted for clarity):
\begin{align}
    \dot{N} & = \dot{N}_{\mr{ev}} + \dot{N}_{\mr{1B}} + \dot{N}_{\mr{3B}} \\
    \dot{E} & = \dot{E}_{\mr{ev}} + \dot{E}_{\mr{1B}} + \dot{E}_{\mr{3B}} + \dot{E}_{\mr{th}}.
\end{align}
The first term on the right-hand side accounts for the effects of free evaporation; the second, for one-body losses due to background collisions; the third, for three-body losses. $\dot{E}_{\mr{th}}$ describes the thermalization between the two species and is given by~\cite{mosk2001mixture} 
\begin{equation}
    \dot{E}^{\mr{Na}}_{\mr{th}} = - \dot{E}^{\mr{Cs}}_{\mr{th}} = \Gamma_{\mr{coll}} \zeta k_{\mr{B}} (T_{\mr{Cs}} - T_{\mr{Na}}),
\end{equation}
where $\zeta = 4 m_{\mr{Na}} m_{\mr{Cs}}/(m_{\mr{Na}}+ m_{\mr{Cs}})^2$ and $m_{\mr{Na(Cs)}}$ is the Na (Cs) atomic mass. The interspecies collision rate is given by
\begin{equation}
    \Gamma_{\mr{coll}} = 4 \pi a^2_{\mr{NaCs}}\bar{v}_{\mr{th}} N_{\mr{Na}} N_{\mr{Cs}} \left( \frac{2\pi k_{\mr{B}} T_{\mr{Na}}}{m_{\mr{Na}}\bar{\omega}_{\mr{Na}}^2} + \frac{2\pi k_{\mr{B}} T_{\mr{Cs}}}{m_{\mr{Cs}} \bar{\omega}_{\mr{Cs}}^2} \right)^{-3/2},
    \label{eq:collision_rate}
\end{equation}
where $k_\mathrm{B}$ is the Boltzmann constant, $T$ is the temperature, $\bar{\omega}=(\omega_x \omega_y \omega_z)^\frac{1}{3}$ is the geometric mean trap frequency, and $\bar{v}_{\mr{th}}$ is the relative thermal velocity given by 
\begin{equation}
    \bar{v}_{\mr{th}} = \sqrt{\frac{8 k_{\mr{B}}}{\pi} \left(\frac{T_{\mr{Na}}}{m_{\mr{Na}}}+ \frac{T_{\mr{Cs}}}{m_{\mr{Cs}}} \right)}.
\end{equation}

We describe the position and momentum distribution of each atomic species in a trapping potential $U(\mathbf{r})$ with trap depth $U_{\mr{min}}$ by a truncated Boltzmann distribution 

\begin{equation}
\begin{split}
    f(\mathbf{r}, \mathbf{p}) & = n_0 \lambda_{\mr{dB}}^3 \exp [-(U(\mathbf{r})+\mathbf{p}^2/2m)/k_{\mr{B}} T]\\
     & \times  \Theta(\eta k_{\mr{B}} T - U(\mathbf{r}) - \mathbf{p}^2/2m),
\end{split}
\end{equation}
where $n_0$ is the peak density, $\lambda_{\mr{dB}} = \sqrt{2 \pi \hbar^2/(m k_{\mr{B}})}$ is the thermal deBroglie wavelength, and $\eta = U_{\mr{min}}/k_{\mr{B}} T$ is the truncation parameter. In the deep trap limit, with large truncation parameter ($\eta > 6$), free evaporation is negligible \cite{luiten1996kinetic}. $U(\mathbf{r}) = m \bar{\omega}^2 \mathbf{r}^2 / 2$ for a dipole trap in the deep trap limit. We obtain the spatial density, integrating over all momentum states as
\begin{equation}
    n(\mathbf{r}) = \frac{1}{(2\pi\hbar)^3} \int f(\mathbf{r},\mathbf{p}) d^3 \mathbf{p} = n_0 \exp (-U(\mathbf{r})/k_{\mr{B}} T),
\end{equation}
and the energy density (neglecting interactions) as
\begin{equation}
\begin{split}
    e(\mathbf{r}) & = \frac{1}{(2\pi\hbar)^3} \int \left(\frac{\mathbf{p}^2}{2m}+U(\mathbf{r} \right) f(\mathbf{r},\mathbf{p}) d^3 \mathbf{p} \\
    & = \frac{3}{2} n_0 k_{\mr{B}} T \exp (-U(\mathbf{r})/k_{\mr{B}} T) + U(\mathbf{r}) n(\mathbf{r}).
\end{split}
\end{equation}
The total atom number $N$ and energy $E= 3N k_{\mr{B}} T$ can be obtained by integrating over position space.

We approximate the evaporation by $\dot{N}_{\mr{ev}} = (\eta - 4) e^{-\eta} \Gamma_{\mr{el}}$ in the deep trap limit~\cite{luiten1996kinetic}, with $\Gamma_{\mr{el}} = n_0 \sigma \bar{v}  / 2\sqrt{2} $ as the intra-species elastic collision rate. $\sigma = 8\pi a_s^2$ is the elastic cross section for identical bosons, $a_s$ is the $s$-wave scattering length, and $\bar{v}=4\sqrt{k_\mr{B} T/\pi m}$. The energy evolution due to evaporation is given by $\dot{E}_{\mr{ev}} = \dot{N}_{\mr{ev}} (\eta + \kappa) k_{\mr{B}} T$, where $\kappa \approx (\eta - 5)/(\eta - 4)$ for a 3D harmonic trap in the deep trap limit \cite{luiten1996kinetic}. The contributions due to inelastic losses are obtained via
\begin{align}
    \dot{N}_{\mr{1B}} + \dot{N}_{\mr{3B}} & = -\Gamma_{\mr{1B}} \int n(\mathbf{r}) d^3 \mathbf{r} - L_{\mr{3B}} \int n(\mathbf{r})^3 d^3r \\
    \dot{E}_{\mr{1B}} + \dot{E}_{\mr{3B}} & = \Gamma_{\mr{1B}} \int e(\mathbf{r}) d^3 \mathbf{r} - L_{\mr{3B}} \int n(\mathbf{r})^2 e(\mathbf{r}) d^3r.
\end{align}
Given the $a_s^4$ scaling of three-body loss coefficients~\cite{weber2003three}, we assume that interspecies three-body losses of the types Na-Na-Cs and Na-Cs-Cs are negligible compared to Na-Na-Na and Cs-Cs-Cs three-body losses. Additionally, in the regimes we consider, Na-Na-Na three-body loss is negligible on our timescales. $\Gamma_{\mr{1B}}$ is the one-body loss rate due technical noise of the ODT. For our experiment, we determine $\Gamma_{\mr{1B}} \approx 0.02$~s$^{-1}$ from the lifetime measurements.

\end{document}